\documentclass[epj,twocolumn]{webofc}

\usepackage[varg]{txfonts}   % Web of Conferences font
\usepackage{graphicx}
\usepackage{tabulary}
\usepackage[exponent-product=\cdot]{siunitx}
\usepackage{amsmath}
\usepackage{aas_macros}

\woctitle{The Innermost Regions of Relativistic Jets and Their Magnetic Fields}
\begin{document}
\newcommand{\be}{\begin{equation}}
\newcommand{\ee}{\end{equation}}
\newcommand{\diff}{\ \mathrm{d}}
\newcommand{\sdiff}{\mathrm{d}}
\newcommand{\del}{\partial}
\newcommand{\unit}{\ \mathrm}
\newcommand{\unitb}{\mathrm}
\renewcommand{\vec}{\mathbf}
\newcommand*{\thead}[1]{\multicolumn{1}{c}{#1}}
\title{The radio morphology of a spatially resolved SSC model}

\author{Stephan Richter\inst{1}\fnsep\thanks{\email{srichter@astro.uni-wuerzburg.de}} \and
        Felix Spanier\inst{2}\fnsep\thanks{\email{felix@fspanier.de}} 
}

\institute{Lehrstuhl f\"ur Astronomie, Universit\"at W\"urzburg, Emil-Fischer-Stra\ss e 31, D-97074 W\"urzburg, Germany
\and
           Centre for Space Research, North-West University, 2520 Potchefstroom, South Africa
          }

\abstract{%
  One of the main, unresolved questions about the nature of quasars is the position of the acceleration site responsible for the highest energies. The attempt to investigate this question in the energy regime with the highest resolution, the radio band, has the downside that no theoretical model exists that can connect these two regimes. The model in this work tries to shrink this gap by extending the general synchrotron self Compton (SSC) model up to length scales in the order of the resolution of radio observations. The resulting spectral energy distributions (SED) show a qualitative improvement in the representation of the radio spectrum. Furthermore the obtained emission morphology shows similar properties to the radio structures observed in jets of quasars. A complete and quantitative connection will however need either much higher numerical effort or an improved methodology.
}
\maketitle
\section{Introduction}
\label{sec:introduction}
In recent years a diverse set of methods for probing as well as modeling of jets in extragalactic sources has been developed. These methods usually focus on a particular region of either space or energy. Furthermore the different timescales on which these sources change need to be analyzed.

The most detailed view of jets, from \SI{}{kpc} to sub-\SI{}{pc} scales, can be obtained from radio interferometry with very large baselines (VLBI)~\cite{mueller_sub-pc}. However, the most inner region between the black hole and the radio-core, where the very high energy (VHE) emission is likely to originate from, is not accessible due to self absorption. In the energies above the radio on the other hand, the central engine together with the jet appears as a point source. In contrast to this fact, modeling of the complete spectral energy distribution (SED) above the radio, is very successful by assuming a region of the size in the order of \SI{E-3}{pc}. This even holds for very dense energy sampling resulting from multi wavelength observations~\cite{mrk501_multi_observation,mrk421_multi_observation}. Especially high peaked BL-Lac sources (HBLs), which show indications of strong Doppler boosting, also show very short variability timescales, supporting the assumption of a very small acceleration and emission region~\cite{PKSflare,Pichel2011a}.

The usually employed synchrotron self Compton models (SSC)~\cite{celotti_first_ssc} are homogeneous in space. Therefore they can not produce radio spectra for the regions and length scales that are observed. On the other hand, large scale jet models depend on (general relativistic) magnetohydrodynamic (MHD) methods. On the one hand, the implementation of acceleration processes is not possible in these models. Hence only synthetic spectra can be used to produce the radiative outputs~\cite{mhd_synchrotron1,mhd_synchrotron2}. On the other hand they are scale invariant, so a motivation for a small region that dominates the VHE emission, in order to explain the observed variability, is challenging.

A connection between these distinct regimes could be obtained in the time domain in the form of correlations between variability patterns of the radio fluxes and morphologies and fluxes in bands of higher energies~\cite{Marscher2008,Leon-Tavares2011a,orienti_gamma_radio_correlation}. However, a theoretical description of this connection does not exist so far.

In section \ref{sec:model} we present a spatially resolved SSC model. It is able to produce non-thermal particle distributions in a small region (i.e. close to a shock) and track it far downstream up to parsec scales. We show in section \ref{sec:results}, that this enables both, the interpretation of radio spectra in terms of SSC models, as well as the connection between the region of VHE emission and the radio blobs of VLBI observations. Section \ref{sec:discussion} will summarize our results. In addition, numerical and conceptional limits of the current model as well as future work will be discussed.

\section{Model}
\label{sec:model}
The SSC paradigm is now well established to be responsible for the double humped SED of at least HBLs. However, the shape of the electron distribution is not determined by this paradigm. The power-law form of the SED and the cosmic-ray spectrum suggests to use a broken power-law for the source intrinsic particle spectrum, too. A time dependent treatment, necessary to constrain the model against observed light-curves, should then incorporate a physical sound acceleration mechanism in which the resulting acceleration timescales depend on physical parameters (see e.g.~\cite{weidinger01}). Moreover, in a spatially resolved model the morphology of the acceleration region has to be well defined.

The most rapid process we know that can produce a power-law distribution is the so called \textit{Fermi~I} acceleration, see e.g.~\cite{Protheroe2004}. This process is based on elastic particle scattering in the vicinity of a MHD shock. Acceleration arises then during the isotropization of the particle distribution after the shock crossing. The scattering processes driving this isotropization will take place at various distances from the shock. Hence this process can be accommodated in an inhomogeneous model quite naturally. Any such model including this process should then mimic the Fermi acceleration as detailed as possible to obtain a connection between the variability patterns of the source and the shock parameters.

The here presented model therefore directly connects the shock with the geometry and its discretization. The pitch angle scattering, driving the particle distribution towards isotropy in the rest frame of the ambient plasma, is included stochastically. Since the particles are also advecting through the system and cool simultaneously, this results in a ratio between acceleration and cooling time that depends on the distance to the shock.

\subsection{Geometry}
\begin{figure}[ht]
  \centering
  \includegraphics[width=0.5\textwidth]{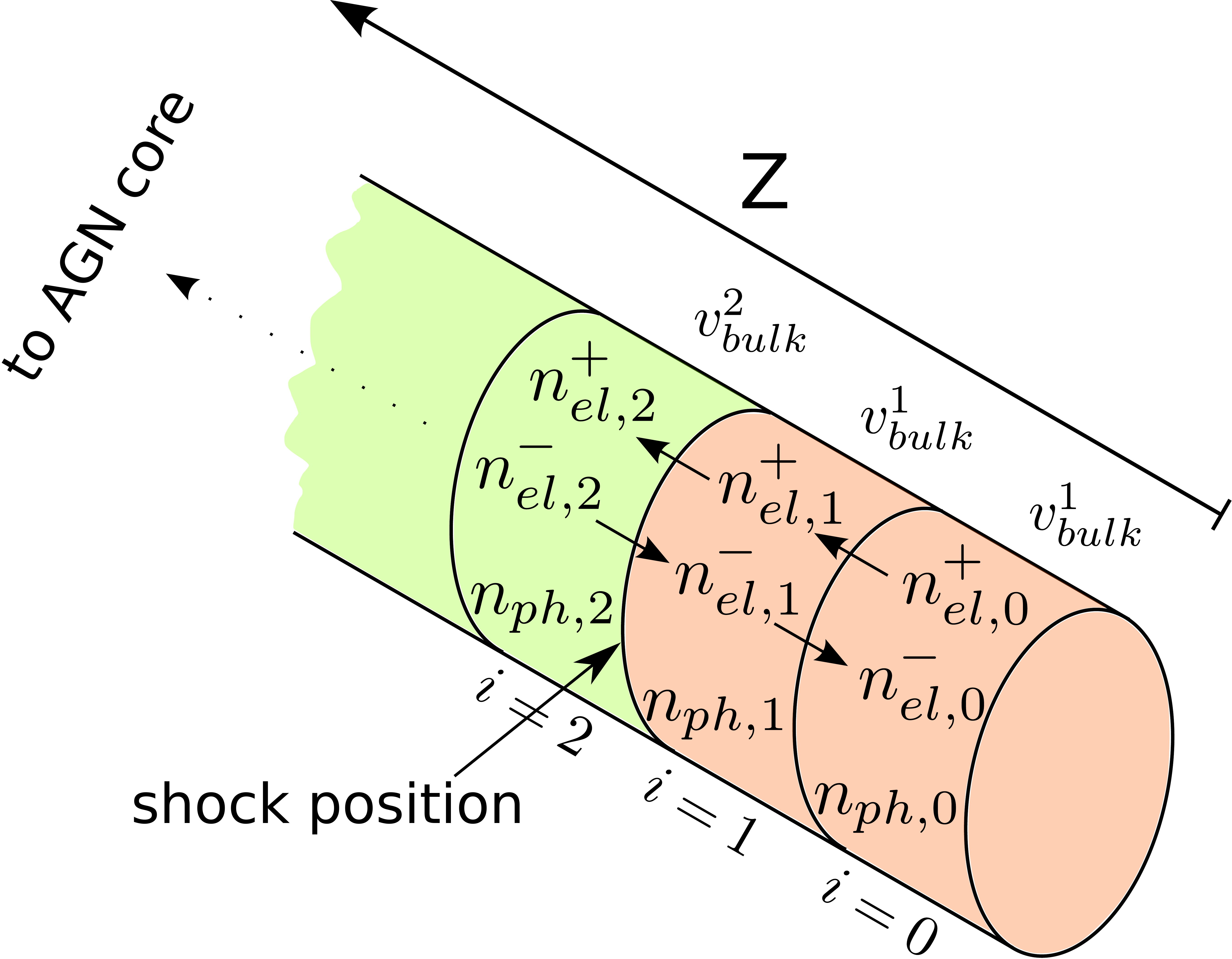}
  \caption{A schematic illustration of the used geometry. The spatial discretization (index $i$) is along the direction of the magnetic field and parallel to the shock normal. A shock is represented as a jump in velocity of the background plasma.}
  \label{fig:geometrie}
\end{figure}

In this model no test particles are included and only the positions of the gyro centers are computed. These two points limit us to non-relativistic and non-oblique shocks. Consequently, the discretization is carried out along the magnetic field lines, parallel to the shock normal. This is shown for the region around the shock in Fig.~\ref{fig:geometrie}.

The resulting slices are indexed with $i$. At each position $z_i$ all parameters can in principle be set freely. However, the $z$ dependence of any parameter should be physically motivated. One can then represent the shock by a jump in velocity of the ambient plasma. From the properties of the shock, its velocity $V_S$ and compression ratio $R$, the bulk velocity in each cell is calculated. The downstream velocity behind each shock $V_P$ (in units of $c$), expressed in the upstream frame is
\be
  V_P=\frac{V_S(R-1)}{R-V_S^2}\quad.
\ee
From here one can, also for multiple shock scenarios, calculate the bulk velocities for each slice in the frame of the first shock. All other calculations then take place in this frame of reference.

Since the \textit{Fermi~I} process depends on the pitch angle scattering, we can not restrict ourselves to the isotropic approximation. On the other hand, a full discretization of the pitch angle $\mu=\cos(\theta)$ (where $\theta$ is the angle between particle momentum and magnetic field) is numerically expensive. Since the pitch angle scattering will produce an isotropic particle distribution on short timescales, it is sufficient to introduce a very crude discretization in the form of two half-spheres:
\be
  n^+(z,\gamma)=\int_{0}^{1}\!n(z,\gamma,\mu)\diff\mu\qquad n^-(z,\gamma)=\int_{-1}^{0}\!n(z,\gamma,\mu)\diff\mu
\ee
Hence the ratio between $n^+$ and $n^-$ is a measure for the anisotropy of the distribution. The introduction of a scattering rate between these two quantities (in the rest frame of the bulk plasma), together with the advection between cells, will lead to Fermi-I acceleration. More precisely, pitch angle scattering between uneven distributions will, on average, lead to particle acceleration.

Using the approximation of isotropy per half-space and the particle speed to be $V_{part}=1$ (since $\gamma\gg1$) one can compute the average advection speed of the particles, only depending on the bulk plasma in the shock frame $\widetilde V_P$:
\be
  V_{adv}^+=\int_{0}^{1}\!\frac{\widetilde V_P+\mu}{1+\widetilde V_P\cdot\mu}\diff\mu
\ee
Changing the integration boundaries to $[-1;0]$ will yield $V_{adv}^-$.

\subsection{Kinetic Equations}
All other processes that will influence the energy distribution of the particles are included in the kinetic equation. It is solved time dependent in each cell $i$ and can be derived from the Fokker-Planck equation. Therefor an integration over $\mu$ has to be performed. Separating the occurring advection terms yields:
\begin{multline}
  \label{eq:gamma_kinetic_el}
  \frac{\del n(\gamma)}{\del t}=\frac{\del}{\del \gamma}\Biggl[D\gamma^2\cdot\frac{\del n(\gamma)}{\del\gamma} \\
  +(\beta_s\gamma^2-2D\gamma+P_{IC}(\gamma))\cdot n(\gamma)\Biggr]+S(z,\gamma,t)
\end{multline}
The time evolution of all $n_i^\pm$ is computed via the above equation. The processes included in Eq.~\ref{eq:gamma_kinetic_el} are the Fermi-II acceleration, synchrotron losses and inverse Compton (IC) scattering. The momentum diffusion coefficient $D=\frac{v_A^2}{9\kappa_\parallel}$ depends on the Alfv\' en speed $v_A$ and the parallel diffusion coefficient $\kappa_\parallel$~\cite{webb83}.
The synchrotron losses are calculated via
\be
	\label{eq:synclosses}
	P_{sync}=\frac{1}{6\pi}\frac{\sigma_TB^2}{mc}\gamma^2=\beta_s\gamma^2\quad,
\ee
where $\sigma_T$ is the Thomsosn cross section~\cite{ginzburg65}.
The losses due to the inverse Compton process are calculated with the full Klein-Nishina cross section~\cite{Blumenthal1970}:
\be
  \label{eq:pic}
  P_{IC,\gamma}=\frac{1}{mc^2}\int\sdiff\nu'\ h\nu'\int\sdiff\nu\ n(\nu)\frac{\sdiff N_{\gamma,\nu}}{\sdiff\nu'\diff t}
\ee
The source function $S$ represents the injection at the boundaries of the simulation box. For simplicity, only delta like (in energy) injections at the upstream edge are used.

The inverse Compton term (Eq.~\ref{eq:pic}) includes the photon density. Due to this back reaction of the produced photon field a non linear system is created and it becomes necessary to calculate the photon density simultaneously. The time dependent equation for $N(\nu,t)$ is
\be
  \label{kinetic_ph}
  \frac{\del N}{\del t}=-c\cdot\kappa_{\nu,SSA}\cdot N+\frac{4\pi}{h\nu}\cdot(\epsilon_{\nu,IC}+\epsilon_{\nu,sync})-\frac{N}{t_{esc}}\quad.
\ee
The dominant gain $\epsilon_{\nu,sync}$ due to synchrotron emission is computed in the Melrose-approximation~\cite{melrose1983}:
\be
  \label{eq:melrose}
  P_\nu(\gamma,\nu)\approx1{,}8\frac{\sqrt3\ q^3B}{m\ c^2}\cdot\left(\frac{\nu}{\nu_c}\right)^\frac{1}{3}\cdot e^{-\frac{\nu}{\nu_c}}
\ee
In the same way the synchrotron self absorption coefficient $\kappa_{\nu,SSA}$ is obtained.
Similar to Eq.~\ref{eq:pic} the photon gains and losses due to the IC process are calculated:
\be
  \label{eq:icwins}
  \epsilon_\nu=\frac{h\nu}{4\pi}\int\sdiff\gamma\ n(\gamma)\int\sdiff\nu'\left(\frac{\sdiff N_{\gamma,\nu'}}{\sdiff\nu\diff t}\cdot n(\nu')-\frac{\sdiff N_{\gamma,\nu}}{\sdiff\nu'\diff t}\cdot n(\nu)\right)
\ee
The catastrophic loss is parameterized by the escape timescale $t_{esc}$.

The total SED is then calculated similar to the model of Blandford and K\"onigl~\cite{Blandford1979}, taking into account light travel times and time dilation.

\section{Results}
\label{sec:results}
The spatial resolution of our model offers the possibility to study the emission from different distances from the the shock. The shape of the SED, especially in the radio regime, is closely related to this, as we will show in section~\ref{sec:emissionmorph}. Before, in section~\ref{sec:radiospectralindex}, the influence of different parameters on the shape of the SED is studied, while in section~\ref{sec:adex} we focus on the adiabatic expansion of the flow behind the shock.
\subsection{Radio spectral index}
\label{sec:radiospectralindex}
\begin{figure}[ht]
  \centering
  \includegraphics[width=0.5\textwidth]{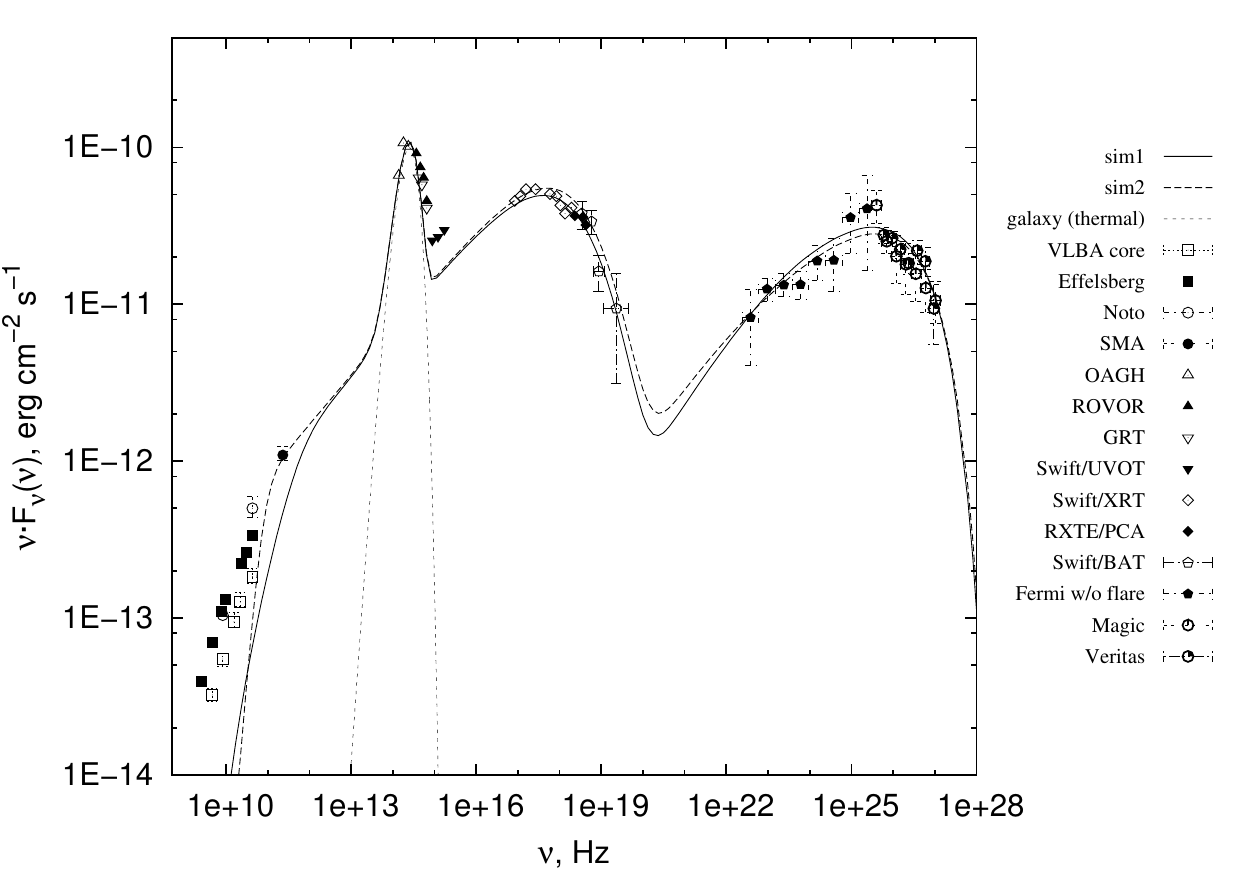}
  \caption{SED obtained during the multi-frequency-campaign in 2009. The fits shown are obtained from the model presented in section~\ref{sec:model} and represent the best fits for high (\textit{sim1}) and low (\textit{sim2}) injection energies, respectively.}
  \label{fig:cc_sed}
\end{figure}

The data basis we are starting from is the multi-frequency-campaign of \textit{Mkn501} in 2009~\cite{mrk501_multi_observation}. This ensures that the qualitativ study will be performed in a parameter space that is physically relevant. Furthermore a comparison to homogeneous models is possibel. In Fig.~\ref{fig:cc_sed} two fits are shown. The fitting was performed ``by eye''. The parameters obtained from these fits are summarized in table~\ref{tab:unicorn_parameters}. They show good agreement with the values found in e.g.~\cite{mrk501_multi_observation}.
\begin{table}[h]
	\centering
	\caption{Parameters obtained from the fits in Fig.~\ref{fig:cc_sed} produced with our spatially resolved model.}
	\begin{tabular}{crrrrr}
		\hline\hline
		 & \thead{$z$} & \thead{$B$} & \thead{$N_{inj}$} & \thead{$\Gamma$} & \thead{$\gamma_{inj}$}\\
		\hline
		\\[-3mm]
		sim1 & $\SI{6E15}{cm}$ & $\SI{0.025}{G}$ & $\SI{2.4E42}{s^{-1}}$ & $34$ & $400$ \\
		sim2 & $\SI{6E15}{cm}$ & $\SI{0.025}{G}$ & $\SI{4.0E43}{s^{-1}}$ & $36$ & $50$ \\
	\end{tabular}
	\label{tab:unicorn_parameters}
\end{table}

In SSC fits, the radio spectrum is usually ignored. This is due to the fact, that the region observed in the radio is much larger than the one assumed for the higher energies. However, often the radio data is included in such fits, by using a very high injection energy or, in stationary models, a large $\gamma_{min}$. The difference between the spectral indeces of the radio in \textit{sim1} and \textit{sim2} is quite obvious. On the one hand side, this raises the question of the origin of these pre-accelerated particles. On the other hand side, the assumption of a minimum Lorentz factor in a region larger than \SI{E15}{cm} would require constant re-acceleration. Otherwise the energies in the particle distribution between $\gamma=1$ and $\gamma_{min}$ would be populated by synchrotron cooling, when moving away from the shock. Without a physical boundary condition for the emission region, the particles will escape from the shock and cooling will eventually dominate the radio emission and steepen the spectrum due to synchrotron-self-absorption.
\subsection{Adiabatic expansion}
\label{sec:adex}
\begin{figure}[ht]
  \centering
  \includegraphics[width=0.5\textwidth]{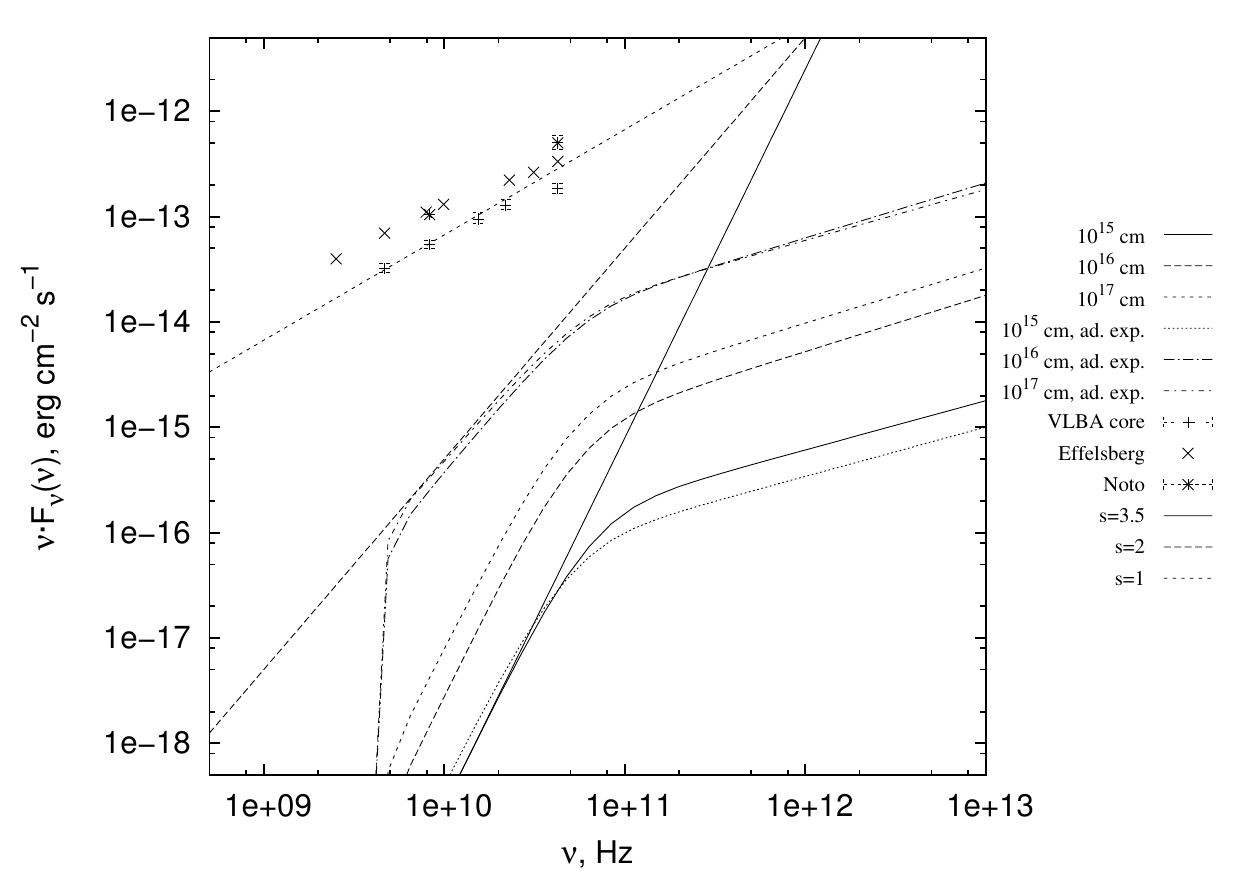}
  \caption{The radio part of the SED for various simulations of different sizes $z$ and opening angles $\alpha$. Shown are the runs with $\alpha=0$ (no adiabatic expansion) and $\alpha=0.5$. The plotted power-laws reflect the self absorbed limit ($s=3.5$), the measured spectral index ($s=1$) and the minimum obtained by our simulations ($s=2$). The total fluxes are scaled for the sake of readability.}
  \label{fig:sed_ad}
\end{figure}

As a more natural way of explaining the observed flat radio spectra we propose an extension of the inhomogeneous SSC model up to the VLBI range ($\SI{E18}{cm}$), including the adiabatic cooling of the SSC particle distribution. The simplest geometry is a linear increase of the radius $R(z)$ behind the shock. In addition to the dilution of the particle density the magnetic field will decrease and the particles will undergo adiabatic cooling:
\be
	B(z)=B_0\left(\frac{R_0}{R(z)}\right)^m\qquad P_{ad}=\frac{1}{3}\frac{\dot V}{V}\gamma=\frac{2}{3}\frac{\widetilde V_P\alpha}{R(z)}\gamma
\ee
Here the zero subscript indicates the values at the shock-position and $\alpha$ denotes the opening angle of the flow. Since we assume a parallel shock and a flow along the direction of the magnetic field we choose $m=2$. The radio part of the resulting SEDs is shown in Fig.~\ref{fig:sed_ad}.
A connection between the size of the simulated region and the spectral index is clearly visible in the case of an adiabatically expanding flow. Although the observed data can't be modeled, this is not expected because of the two orders of magnitude that still separate the sizes of the observed region and the simulation. It can also be observed, that in the current setup an extension beyond a size of $z=\SI{E16}{cm}$ only showed marginal improvement. In Fig.~\ref{fig:sed_ad_morph} of the next section the reason for this behavior will become obvious. Due to the huge numerical costs the question whether this is a systematic effect or is strongly parameter dependent (e.g. the magnetic field structure) has be left to future work at this point.
\subsection{Emission morphology}
\label{sec:emissionmorph}
\begin{figure}
  \centering
  \includegraphics[width=0.5\textwidth]{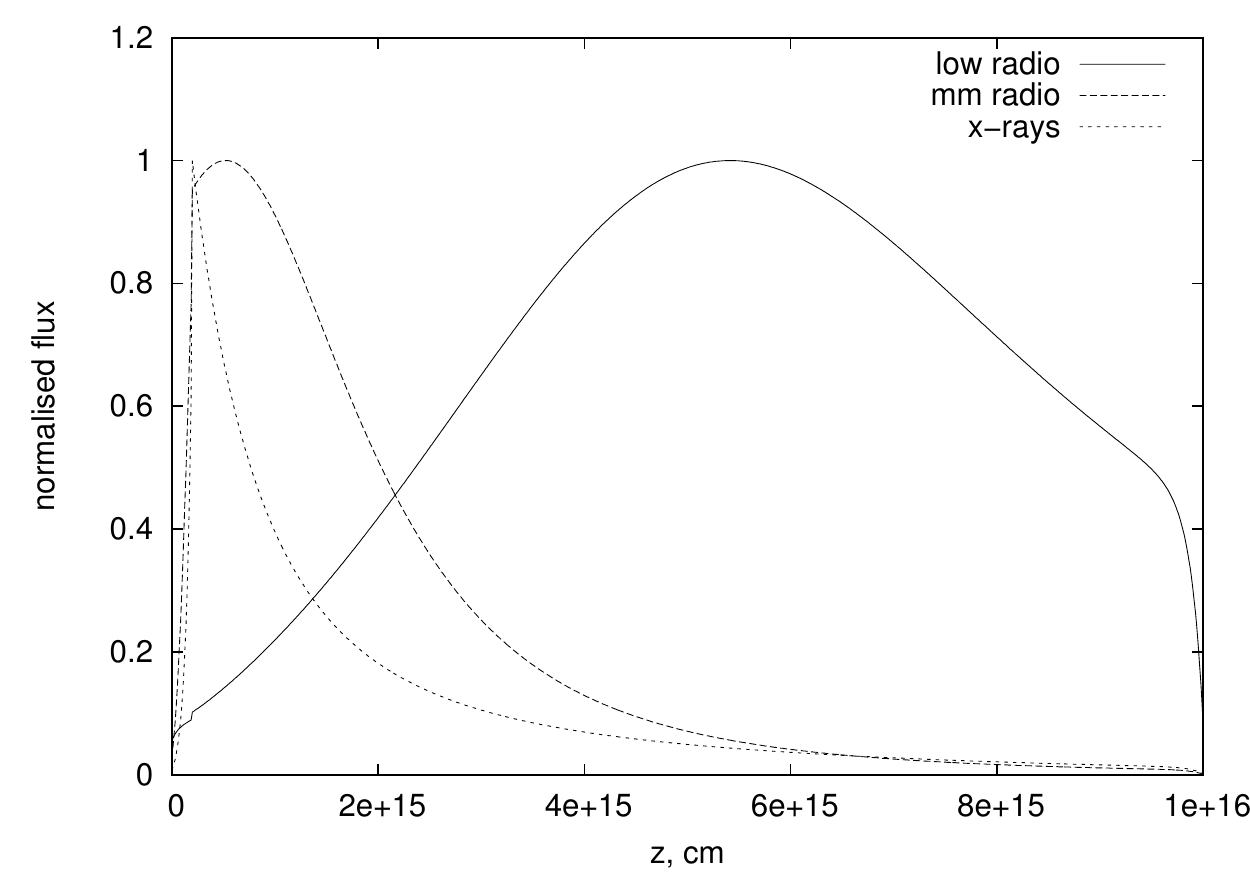}
  \caption{Distribution of the flux over the simulated region for different energy regimes. The shock is located at the very left side, where the highest energies have there peak value.}
  \label{fig:sed_ad_morph}
\end{figure}

The distribution of the emission depending on the distance from the shock is, in principle, directly comparable with VLBI observations. However, resolving four orders of magnitude is numerically quite expensive. Nevertheless one can estimate from Fig.~\ref{fig:sed_ad_morph} a qualitative pattern.

The photon energies produced by the most energetic particles are confined to a region very close to the shock. This is caused by the rapid synchrotron cooling at these energies. Moving to smaller energies will at first delay the cooling decay, until the self absorbed regime is reached. Here, the fluxes close to the shock are negligible. At larger distances however, where smaller densities and magnetic fields are present, smaller and smaller energies are no longer self absorbed. This results in, with increasing wavelength, growing structures at larger distances from the shock.

In the simulations so far most parts of these structures, even for small radio energies, were already captured within $z\approx\SI{E16}{cm}$. Consequently, the simulation one order of magnitude larger won't show significant differences in Fig.~\ref{fig:sed_ad}. However, the emerging pattern fits qualitatively well to the core shift (with respect to the wavelength) usually observed in AGN.

\section{Discussion}
\label{sec:discussion}
The numerical model presented in this work extends the canonical SSC model to spatial scales much larger than the region that has to be resolved in order to explain the observed fast variability of the high energy emission.
The tracking of the accelerated particle distributions far downstream in different physical environments leads to the conclusion, that a connection between the VHE and radio regime are most likely possible under the assumption of an adiabatically expanding flow.
First results presented in this work show an improvement of the fit in the self absorbed regime, when increasing the size of the simulation box up to $\SI{E17}{cm}$.
Simultaneously the emerging emission morphology can qualitatively represent radio structures generated from VLBI observations.
The quantitative dependence of the emerging structures will be studied in future work.

Furthermore, the time correlation between different energy bands usually yield typical time lags between the highest energies and the radio regime.
In Fig.~\ref{fig:sed_ad_morph} there is clearly a potential for time lags between the prompt fluctuations around the shock and the delayed and smeared out variability of the radio structures far downstream.
This in turn would allow a prediction for the shock position - and hence the origin of the gamma rays - relative to the maxima of the radio fluxes at different energies.
Again, quantitative studies of particular flare scenarios are left to future work.

\begin{acknowledgement}
SR wants to thank GK 1147 for their support. 
\end{acknowledgement}
%
% BibTeX or Biber users please use (the style is already called in the class, ensure that the "woc.bst" style is in your local directory)
\bibliography{apj-jour,ref}

\end{document}